# Formation and kinetics of transient metastable states in mixtures under coupled phase ordering and chemical demixing

Ezequiel R. Soulé[1] and Alejandro D. Rey[2]

1. *Institute of Materials Science and Technology (INTEMA), University of Mar del Plata and National Research Council (CONICET), J. B. Justo 4302, 7600 Mar del Plata, Argentina*

2. *Department of Chemical Engineering, McGill University, Montreal, Quebec H3A 2B2, Canada*

**Abstract**

We present theory and simulation of simultaneous chemical demixing and phase ordering in a polymer-liquid crystal mixture in conditions where isotropic-isotropic phase separation is metastable with respect to isotropic-nematic phase transition. It is found that mesophase formation proceeds by a transient metastable phase that surround the ordered phase, and whose lifetime is a function of the ratio of diffusional to orientational mobilities. It is shown that kinetic phase ordering in polymer-mesogen mixtures is analogous to kinetic crystallization in polymer solutions.

Rapid cooling of a liquid may result in a solid whose structure and composition is different when using slow cooling, even in the absence of vitrification, because transient metastable phases may emerge. For example in a mixture, liquid-liquid (LL) equilibrium can be buried below the liquid-solid (LS) transition, and with a fast and large undercooling, the metastable LL phase separation can precede crystallization [1]. The emergence of transient metastable phases and the evolution of phase transformation trough transient metastable states is known as the empirical Ostwald step rule [2]. Accordingly, crystallization consists of a sequence of chemical and structural changes rather than a single-step energy minimization step, known as kinetic crystallization pathway [1,2]. This is particularly relevant to crystallization of polymer solutions [1], where buried metastable states below the crystallization temperature are readily accessible through thermal quenches [3,4].

Traditionally, the main mechanism considered for the emergence of the transient metastable phase was its higher nucleation rate as compared with the stable one. In the last decade, Bechoeffer et al [5,6] described a new dynamic mechanism for the formation of metastable phases, based on the Landau-Ginzburg equation for a non-conserved order parameter (model A [7]). It as found [5,6] that an interface separating the stable high- and low- temperature phases can spontaneously split into two interfaces, one separating the high temperature phase and the metastable phase, and the other separating the metastable and the low temperature phase, independently of nucleation events. For a single order parameter, the necessary and sufficient condition for splitting is that the velocity of the second interface is higher that the velocity of the first. In case of more than one order parameter, the splitting can be hysteretic or need a finite magnitude perturbation, so the splitting may depend also on the initial conditions.

The emergence of metastability by front splitting considered in [5,6] is limited to non-conserved order parameters (NCOP). For mixtures, the conserved concentration must be taken into account. The minimum model for a first order transition in mixtures must take into account one conserved and one non-conserved order parameter (model C [7]). As the dynamics of both order parameters is different, a more complex behaviour appears in these systems. Recently Fischer and Dietrich [8] studied spinodal decomposition for model C and found that depending on the relative values of diffusional and ordering mobilities, different mechanisms and metastable phases could appear; for example, if ordering is much faster than diffusion, the system first becomes ordered at constant concentration, and then a phase separation takes place, whereas in the opposite case phase-separation and phase ordering evolve simultaneously. Simulations [9,10] of polymer-liquid crystal mixtures exhibiting simultaneous phase ordering and chemical demixing have shown that liquid crystal orientational order interacts with chemical demixing producing morphologies sensitive to kinetic factors.

The objective of this letter is (1) to demonstrate that simultaneous demixing and liquid crystal phase ordering follow the principles of Oswald stage rule [1] and that transient metastable states are likely to be found during mesophase formation, thus establishing a connection between kinetic crystallization in polymer solutions [1] and mesophase formation in polymer-liquid crystal solutions, and (2) to show that the multiple metastable fronts in NCOP systems [5,6] also appear in mixed order parameter systems.

As a model system, in this letter we simulate the 1D phase transition in a polymer-mesogen mixture, when the LL phase separation is metastable with respect to isotropic-nematic transition. An already formed nematic/isotropic interface is taken as the initial state and the lifetime of the transient metastable state is established by changing the diffusional to orientational mobility ratio.

The free energy density of the system is the sum of homogeneous and gradient contributions. The Flory-Huggins theory is used for the mixing free energy, in combination with the Maier-Saupe theory for nematic order [9,10,11]. The homogeneous free energy (per mole of cells) is [11]:

$$\frac{f^h}{RT}=\frac{\phi}{r_c}\ln(\phi)+\frac{1-\phi}{r_p}\ln(1-\phi)+\chi\phi(1-\phi)+\frac{3}{4}\frac{\Gamma}{r_c}(1-\phi)^2\,\mathbf{Q}:\mathbf{Q}-\frac{(1-\phi)}{r_c}\ln\left(\frac{I_0}{2}\right) \qquad (1)$$

where

$$I_0=\int_0^{2\pi}\int_0^{\pi}\exp\left[\frac{3}{2}\Gamma(1-\phi)\mathbf{Q}:\left(\boldsymbol{\pi\pi}-\frac{\boldsymbol{\delta}}{3}\right)\sin^2\theta\, d\theta d\omega\right],$$

$\phi$ is the liquid crystal volume fraction, the symmetric and traceless tensor $\mathbf{Q} = S\mathbf{nn} + P(\mathbf{ll} - \mathbf{mm})$ is the quadrupolar order parameter [9], where $S$ and $P$ are the scalar uniaxial and biaxial parameters, **n**, **m** and **l** are the eigenvectors of **Q**, $r_c$ and $r_p$ are the ratios of molar volume of liquid crystal and polymer with respect to the cell volume, $\chi$ is the mixing interaction parameter, $\Gamma$ is a Maier-Saupe interaction parameter (both are functions of $1/T$), $\boldsymbol{\pi}$ is a unit vector and $\boldsymbol{\delta}$ is the identity matrix, $R$ is the gas constant and $T$ is the temperature. $I_0$ was approximated by a polynomial expression in terms of the invariants of **Q** as $I_0=a(\mathbf{Q}:\mathbf{Q})+b\left[(\mathbf{Q}\cdot\mathbf{Q}^T):\mathbf{Q}\right]+c(\mathbf{Q}:\mathbf{Q})^2+d(\mathbf{Q}:\mathbf{Q})\left[(\mathbf{Q}\cdot\mathbf{Q}^T):\mathbf{Q}\right]$. The coefficients $a$, $b$, $c$ and $d$ were obtained from a least-squares fitting of the numerical solution of the integral. As the evaluation of a polynomial expression is easier than the numerical solution of the integral, this is beneficial from the simulation point of view. The gradient free energy is given by gradients in concentration and order [9,10]:

$$f^{g} = l_{\phi}\left(\nabla\phi\right)^{2} + l_{Q1}\nabla\mathbf{Q}\vdots\nabla\mathbf{Q} + l_{Q2}\left(\nabla\bullet\mathbf{Q}\right)\bullet\left(\nabla\bullet\mathbf{Q}\right) + l_{Q\phi}\left(\nabla\bullet\mathbf{Q}\right)\bullet\nabla\phi \qquad (2)$$

The time-dependent Ginzburg-Landau formulation and the Cahn-Hilliard equation are used to simulate the time evolution of **Q** and $\phi$ [10]:

$$\frac{\partial\mathbf{Q}}{\partial t} = M_{Q}\left(-\frac{\partial f}{\partial\mathbf{Q}} + \nabla\cdot\frac{\partial f}{\partial\nabla\mathbf{Q}}\right) \qquad (3)$$

$$\frac{\partial\phi}{\partial t} = M_{\phi}\nabla^{2}\left(\frac{\partial f}{\partial\phi} - \nabla\cdot\frac{\partial f}{\partial\nabla\phi}\right) \qquad (4)$$

The numerical methods used to compute phase diagrams are given in [9,10]. Comsol Multiphysics was used to solve eqn.(3,4), with quadratic Lagrange basis functions; standard numerical techniques were used to ensure convergence and stability.

Fig. 1 shows the thermodynamic phase diagram used in this study computed using eqn.(1); see caption. The buried metastable I+I region arises because nematic (N) ordering decreases the energy and moves the common tangent outwards, resulting in a metastable gap buried within the N/I binodal. The issue to be established is how the metastable gap influences the phase ordering-demixing process. The representative initial condition is an interface separating an isotropic phase with $\phi$=0.8 (left dot in Fig.1) from a nematic phase (right dot in Fig.1) with $\phi$ and **Q** given by the equilibrium conditions. Different values of diffusional-to-phase ordering mobility ratio $M_R = M_\phi/l^2 M_{\mathbf{Q}}$ were used, where $l$ is the characteristic length, which was defined as $l=(l_\phi/RT)^{1/2}$. The spatial position is expressed in units of this characteristic length and the time is expressed in units of $\tau$ = $(M_Q RT)^{-1}$.  The gradient parameter ratios $l_{Q1}/l_\phi = l_{Q2}/ l_\phi = 0.1$ and $l_{\phi Q}/l_\phi$ = 0.5 were used in all simulations.

The kinetics of the phase transition will result from the combination of ordering and diffusion kinetics. If the characteristic velocities of both processes are similar the overall kinetics will show an intermediate behaviour. The limiting cases of diffusional control and ordering control can be analyzed with simplified semi-analytical models, as follows. (A) Diffusional kinetics: we can use Fick´s law, which is a sharp-interface equivalent to Cahn-Hilliard equation if the diffusivity is $D = M_\phi\, \partial^2 f/\partial\phi^2$ and the gradient terms (which are expected to be important only in the interface) are neglected in the bulk phases. The assumption of constant mobility implies a non-constant diffusivity, but in order to find an analytical solution it will be assumed to be constant, evaluating the second derivative of the free energy at the average of the concentrations at the interface and at the bulk. The boundary and initial conditions are $\phi|_{x=X_{INT}} = \phi^i{}_{INT}$ ; $\phi|_{x=\infty} = \phi^i{}_{bulk}$ ; $\phi|_{t=0} = \phi^i{}_{bulk}$ ; $(\phi^n{}_{INT} - \phi^i{}_{INT})v = D\partial\phi/\partial x$, where $X_{INT}$ is the position of the interface, $\phi^i{}_{bulk}$ is the concentration in the isotropic bulk phase, $\phi^i{}_{INT}$ and $\phi^n{}_{INT}$ are the concentration in the nematic and isotropic sides of the interface (equilibrium concentrations), and $v$ is the velocity of the interface. The concentration profile can be found using a similarity transformation [11], and is given by:

$$\phi = \frac{\phi^i{}_{bulk} - \phi^i{}_{INT}}{1 - erf(\sigma)}\left[erf(\eta) - erf(\sigma)\right] + \phi^i{}_{INT} \qquad (5)$$

where *erf(x)* is the error function and the factor $\sigma$ satisfies:

$$2\left(\phi^n{}_{INT} - \phi^i{}_{INT}\right)\sigma = \frac{\phi^i{}_{bulk} - \phi^i{}_{INT}}{1 - erf(\sigma)} \exp\left(-\sigma^2\right)$$

The velocity of the interface is $v = \sigma D^{1/2} t^{-1/2}$

(B) Phase Ordering Kinetics: an expression for the velocity can be found following the procedure of [12]. If the process is ordering-controlled the phase ordering front velocity is $v = \frac{1}{\beta}\mathbf{L}$, where $\beta = \int_{\Delta N}^{\Delta I} \partial\mathbf{Q}/\partial x : \partial\mathbf{Q}/\partial x\, dx$ is the interfacial viscosity, $L = \int_{\Delta N}^{\Delta I} \partial f/\partial\mathbf{Q} : \partial\mathbf{Q}/\partial x\, dx$ is the bulk stress load, and $\Delta I$ and $\Delta N$ are the positions of isotropic and nematic boundaries of the interface. It has been observed in the simulations that the drop in **Q** across the interface is more abrupt than the drop in $\phi$, so the integrand in the expression for $L$ is non-zero in a region where $\phi$ varies slightly and it's close to the nematic phase value. We can assume that all the drop in **Q** is located is a region where $\phi$ is constant (this will give the maximum driving force for ordering, which is reasonable considering that the system is controlled by ordering), so we can calculate the bulk stress load as $L = f(\mathbf{Q}^n, \phi^n_{INT}) - f(0, \phi^n_{INT})$. The interfacial viscosity can be found by numerical integration of the order parameter profiles found in the simulations.

The previous derivations show the distinctive characteristics of each type of dynamics: the velocity of a purely diffusional process decreases with time as $v \propto t^{-1/2}$, while for a purely ordering process $v$ is a constant [12]. For a mixed process, an intermediate behaviour is expected. Next we discuss numerical solutions to eqns. (3,4) and use the established velocity scalings to rationalize the computations.

For high values of $M_R$ (diffusion is faster than ordering), the interface was observed to spontaneously split in two, one separating the two metastable isotropic phases, and the second one separating an isotropic phase and the equilibrium nematic phase, as shown in fig. 2a. This is in agreement with [5,6]: the whole interface is considered as being composed by two interfaces: nematic – isotropic 1 and isotropic 1 – isotropic 2. As ordering is slow, the dynamics of the first interface will be controlled by ordering and it will be slow, whereas the second interface has a pure diffusional dynamics and it moves faster, so the interface splits in two. For NCOP [5,6] both interfaces have constant velocity (ordering dynamics), and the distance between the two interfaces remains constant or increases linearly with time. But in the present mixed order parameters case, the I-N interface has a constant velocity but the I-I interface slows down with time, so at short times the interfaces separate but after some time (governed by the mobility ratio) they merge again, as shown in fig. 2.b. Thus, under phase ordering control the metastable phase has a finite lifetime, as in Ostwald step rule [1,2].

When the value of $M_R$ decreases the lifetime of the metastable phase decreases. A theoretical plot of the metastable phase lifetime versus $M_R$ was constructed from the simplified model and compared with simulations. The velocity of the nematic/isotropic interface can be calculated assuming that its kinetics is controlled by ordering, with $L/RT$=0.0162 and $\beta/M_Q l$ = .0663, giving $v_{N-I} = 0.23$. The velocity of the isotropic/isotropic interface can be approximated solving Fick´s law, with $\phi^l_{INT}$ = 0.967 and $\phi^r_{INT}$ = 0.776, giving $v_{I-I} = 3.78 M_R^{1/2} t^{-1/2}$. The time at which both interfaces intersect each other is given by $t = 4.9 \cdot 10^{-3} M_R$. This is plotted in fig. 3, together with the merging times observed in the simulations.

For sufficiently small mobility ratio the merging time will be so small that the maximum separation between the interfaces will become smaller that the interface width, so there will be an “incomplete” splitting. The separation between both interfaces, according to the simplified model, is $\Delta x = 8.11 M_R^{\frac{1}{2}} t^{\frac{1}{2}} - .23t$, and the maximum separation is $\Delta x = 3.25 \cdot 10^{-4} M_R$. For a mobility ratio of $2.4 \cdot 10^4$, the maximum separation is 7.8, and the interface thickness (observed in simulations) is about 20, so “incomplete” splitting is expected. This was observed in simulations, where splitting was not seen, but a shoulder appeared in the interface at short times and then disappeared. This can be seen in fig 4, which is a plot of the concentration profiles across the interface at different times. The merging time in this case was taken as the time at which the shoulder disappears completely.

Having analysed the splitting-merging mechanism, next we establish the post-merging kinetics. The formed single interface is found to present a dynamics that is neither purely ordering nor purely diffusional. Figure 5 shows the results from the simulation with $M_R$=$2.4 \cdot 10^4$ (small merging time – incomplete splitting). We also plot the limiting velocities calculated with the simplified models for ordering- and diffusion-control. It can be seen that, at short times the kinetics is closer to be ordering-controlled, but as time goes on, there is a transition to diffusional control. This is because ordering kinetics is independent of time, while diffusion slows down as the concentration profiles develop, so eventually, at some time the kinetics must become diffusion-controlled.

As shown in figure 2a, the concentration drop at the nematic-isotropic interface when the metastable phase forms is small, in particular it is smaller than when a unique interface is present. This implies a different value of interfacial tension. As texturing and defect dynamics depends on interfacial tension, the formation of a metastable phase is expected to have an impact in texture and defect formation for bi- or tri-dimensional geometries, and thus it will affect the optical properties of the material. Furthermore, this phenomenon is not restricted to polymer-liquid crystal mixtures; it could be present in any system with coupled conserved and non-conserved order parameters (e.g.: metallic alloys). Another implication that arises from our analysis is that, even when the overall kinetics of the process will be diffusion-controlled at long times, the dynamic behaviour is more complex at short or intermediate times, and this can affect the final morphologies as well as the overall time scale of the process.

In conclusion, the dynamics of an isotropic-nematic interface in a system exhibiting simultaneous chemical demixing and liquid crystal phase ordering was simulated with different diffusional-orientational mobility ratios with the objective of establishing that mesophase formation exhibit analogous behaviour as kinetic crystallization of some polymer solutions [1]: (a) a buried metastable state in the phase diagram, (b) transient metastable appear as predicted by the Ostwald rule of kinetic crystallization, and (c) transformations via multiple stable-metastable interfaces arise due to the distinct kinetics of phase ordering and diffusional fronts. For high mobility ratios the interface spontaneously splits in two, one separating the two metastable isotropic phases, controlled by diffusion, and the other one separating one isotropic and the nematic phases, controlled by ordering, in accordance with the mechanism found by [5,6] for non-conserved order parameters. But in the present case, as the dynamics of both interfaces are different, the interfaces merge again after some time, so the metastable phase has a finite lifetime following Ostwald step rule [1,2]. After merging, the interface has a mixed dynamical behavior, nor purely ordering-controlled nor diffusion-controlled. At the beginning the velocity is close to the ordering-controlled velocity, but as diffusion slows down with time, the system transitions to diffusional control and the velocity decreases with time. Finally, it is shown that mesophase formation in multiple component systems, like kinetic crystallization in polymer solutions [1], expands the richness of material transformations already revealed in the long study of polymeric systems .

**Acknowledgement**

ADR acknowledges partial support from the Natural Science an Engineering Research Council of Canada (NSERC) and the Petroleum Research Fund. ERS acknowledges a scholarship provided by the National Research Council of Argentina (CONICET).

**Figure Captions**

**Figure 1.** Computed phase diagram based on eqn. (1); for: $r_c$=2.74, $r_p$ =200, $\chi$=-.645+.18/Γ. Γ is the inverse of the nematic interaction parameter and $\Gamma^{-1}$ is a dimensionless temperature, and ϕ is the volumetric fraction of liquid crystal. The dots indicate the coordinates of both phases used as initial conditions when solving eqns.(3,4).

**Figure 2.** Interface splitting observed with $M_R = 2.4{\cdot}10^6$. (a) Concentration ϕ (full line, left axis) and scalar order parameter $S$ (dashed line, right axis) profiles at $t$ = 650. $X$ represents the dimensionless position (b) Dimensionless position of the two interfaces shown in (a) (dashed line: N-I interface; full line: I-I interface), as a function of dimensionless time.

**Figure 3**. Merging time as a function of mobility ratio, from simulations (squares) and from eqn. 4 (line). As $M_R$ increase the lifetime of metastable state increases.

**Figure 4.** Concentration profiles across the interface for $M_R$=$2.4{\cdot}10^4$ for dimensionless times (curves from left to right): 20, 60, 120, 180. ϕ is the volumetric fraction of liquid crystal and $X$ the dimensionless position. The arrows indicate the location of the shoulder.

**Figure 5.** Dimensionless interfacial velocity as a function of dimensionless time; squares are from simulations, and the dashed lines are calculated with the simplified diffusion and phase ordering models.

**Figure 1**

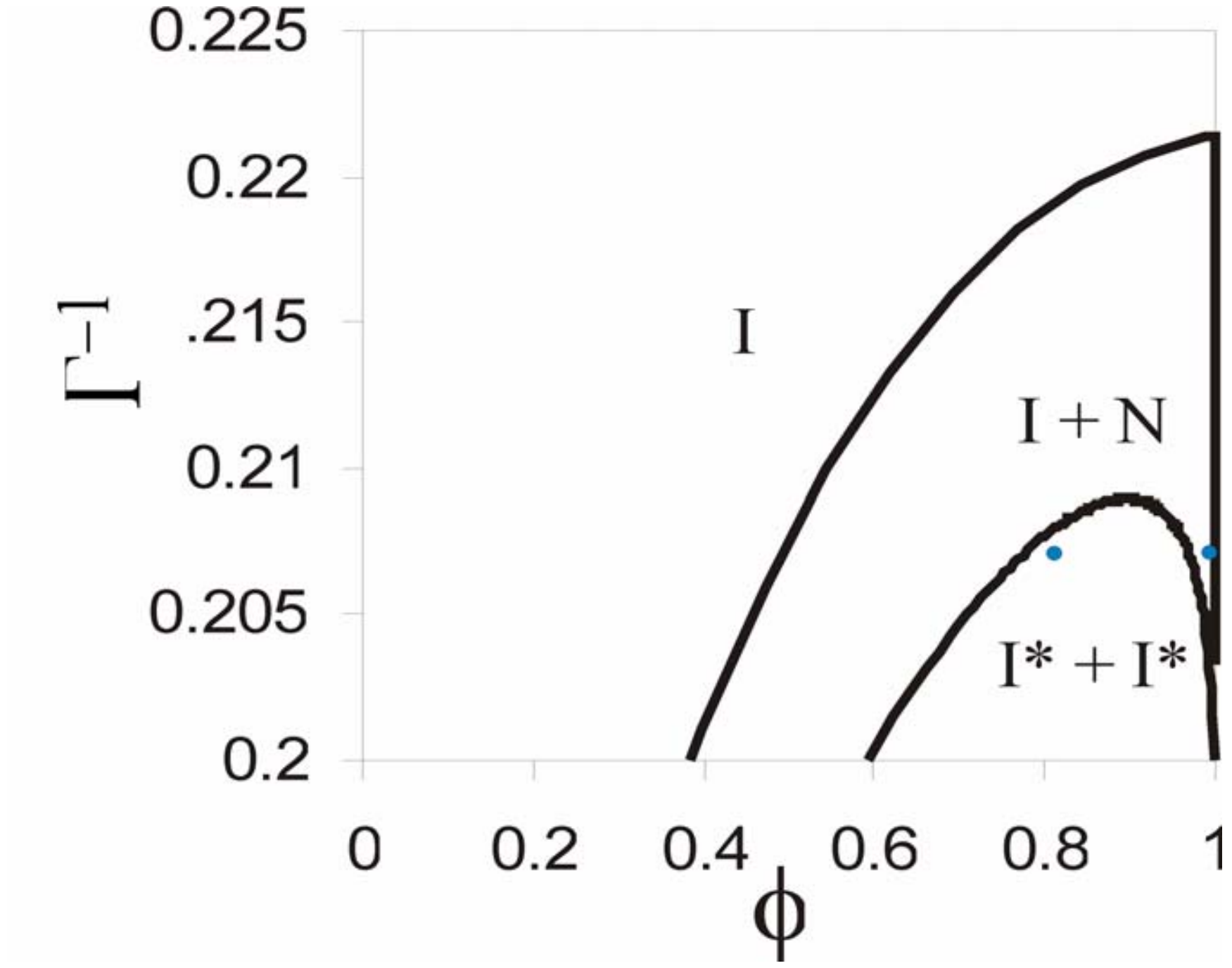

**Figure 2**

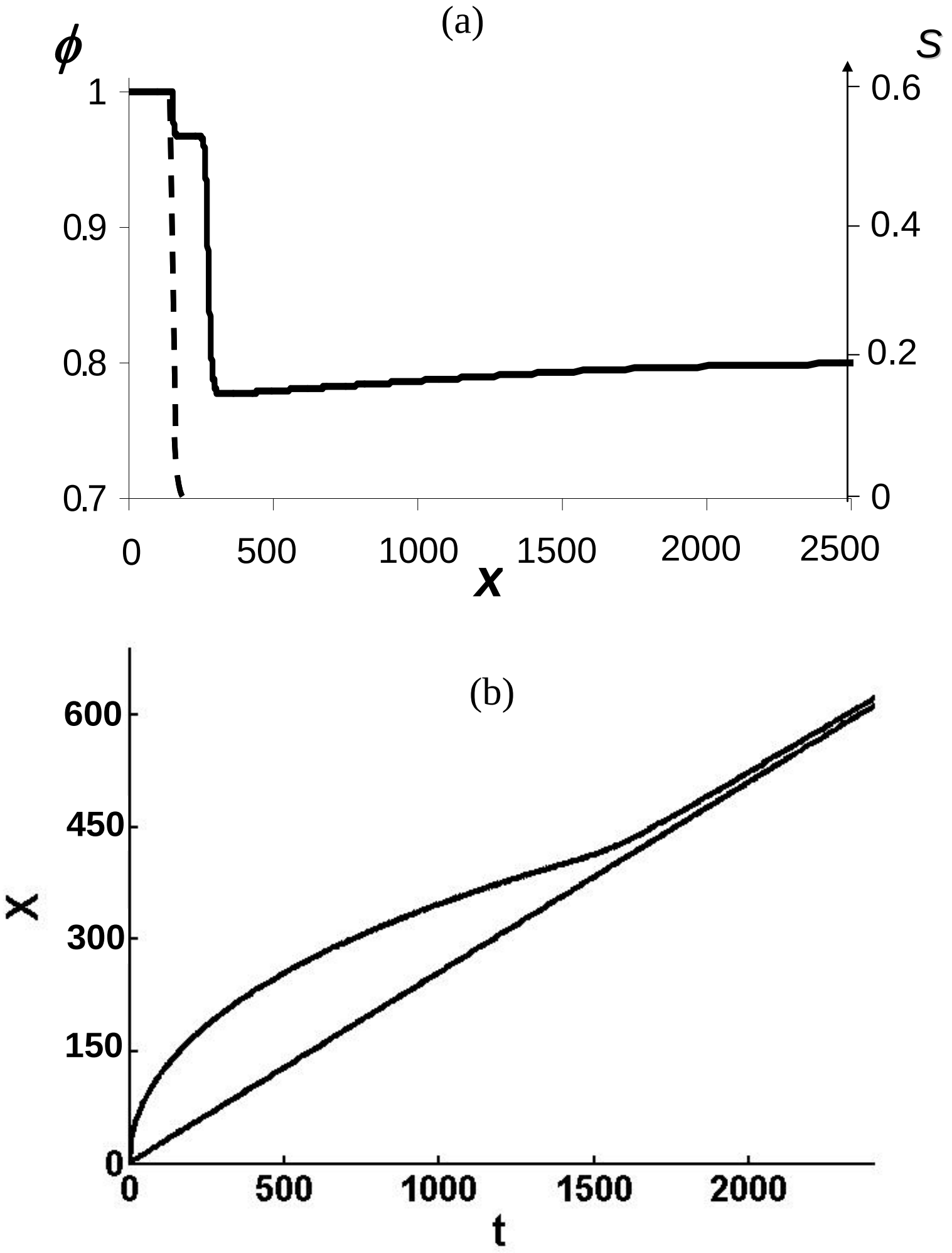

**Figure 3**

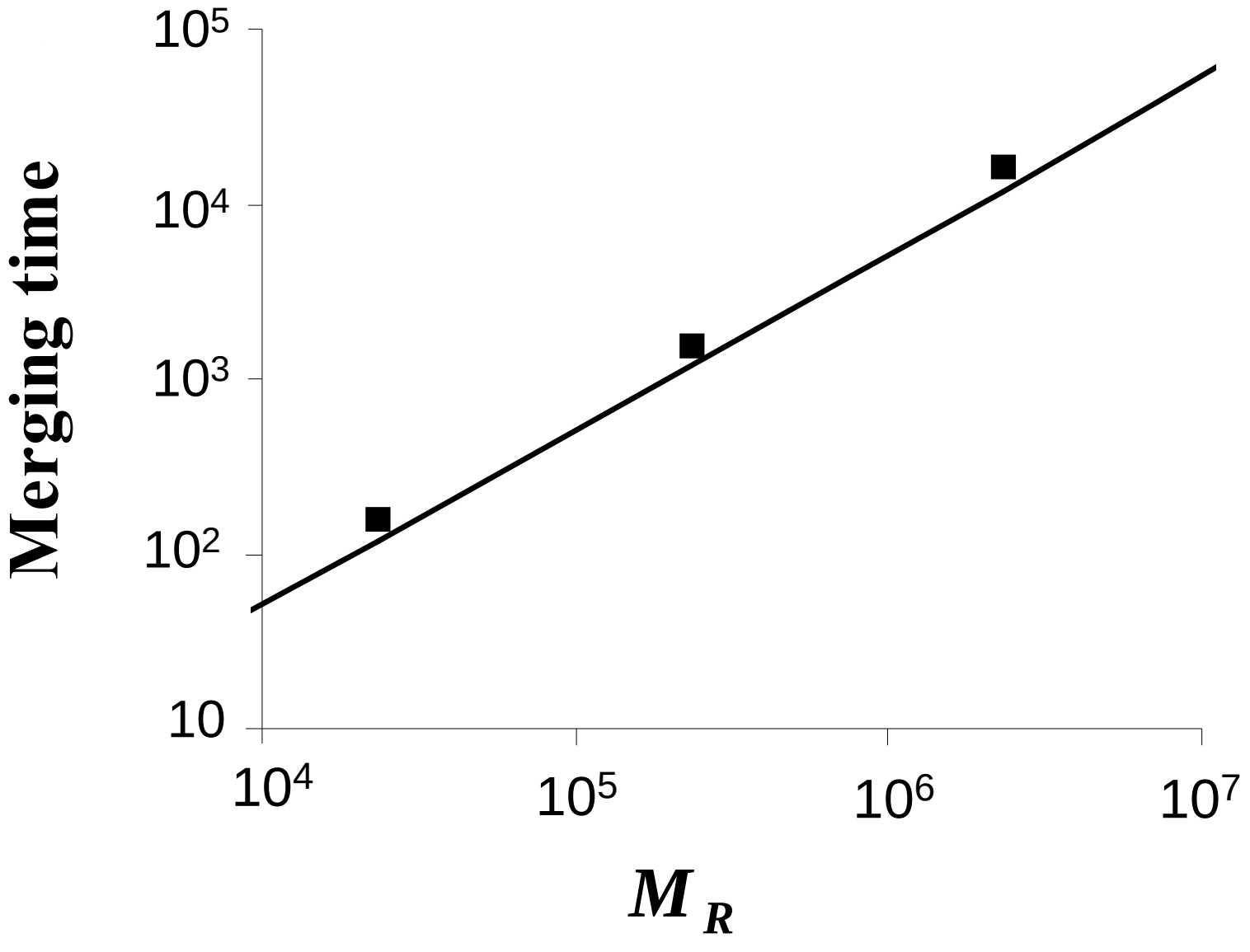

**Figure 4**

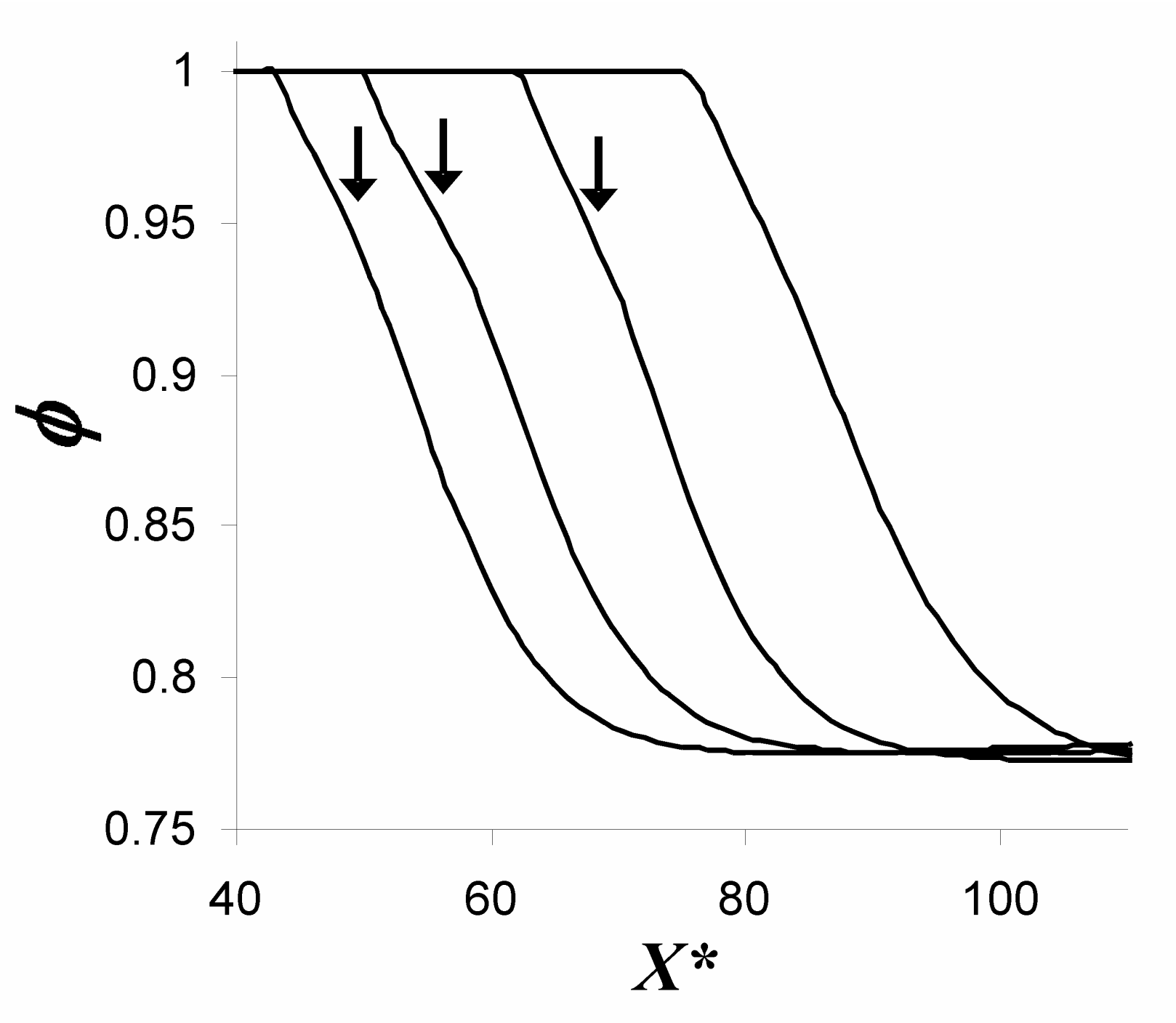

**Figure 5**

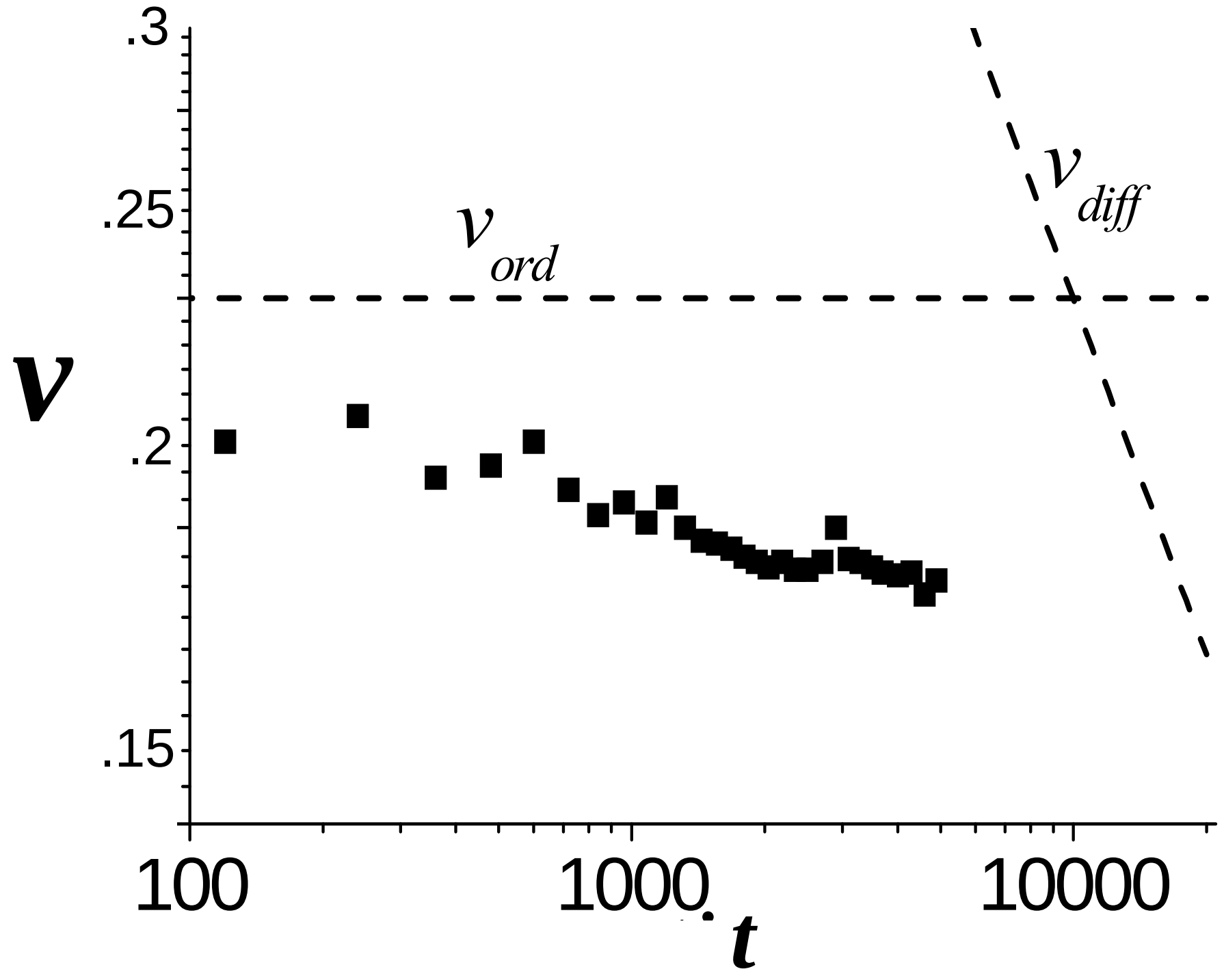